\documentclass[aps, pre, twocolumn, amsmath, superscriptaddress,showkeys,showpacs]{revtex4-1}

\usepackage{fancyhdr}
\usepackage{calc}
\usepackage{amssymb}
\usepackage{setspace}
\usepackage{amsfonts}
\usepackage[innercaption]{sidecap}
\usepackage{graphicx}% Include figure files
\usepackage{graphicx,epstopdf}
\usepackage{xcolor, soul}
\usepackage{dcolumn}% Align table columns on decimal point
\usepackage{bm}% bold math
\usepackage{mathrsfs} % script-like, curvy letters.
\usepackage{amsmath} %curvy letters.
\usepackage[colorlinks=true,linkcolor=blue,citecolor=blue]{hyperref}%
\usepackage{fontenc}
\usepackage{float}
\usepackage{amsthm}
\usepackage{subfigure}
\usepackage{color}
\usepackage{ragged2e}
\usepackage{enumerate}
\usepackage{hyperref}
\begin{document}	
	\renewcommand{\arraystretch}{1.5}
	
	\newcommand{\R} %PD%
	{
		\mathbb R
	}
	
	\newcommand{\M} %PD%
	{
		\mathbb M
	}
	
	\newcommand{\A} %PD%
	{
		\mathcal A
	}

	\newcommand{\eref}[1]{Eq.~(\ref{#1})}%
\newcommand{\Eref}[1]{Equation~(\ref{#1})}%
\newcommand{\fref}[1]{Fig.~\ref{#1}} %
\newcommand{\Fref}[1]{Figure~\ref{#1}}%
\newcommand{\sref}[1]{Sec.~\ref{#1}}%

%	\title{Transient dynamics of deterministic systems under resetting}
\title{Mitigating long transient time in deterministic systems by resetting}
	\author{Arnob Ray}
	\thanks{Equal contribution}
	\affiliation{Physics and Applied Mathematics Unit, Indian Statistical Institute, 203 B. T. Road, Kolkata 700108, India}
	\author{Arnab Pal}
			\thanks{Equal contribution}
			\email{arnabpal@mail.tau.ac.il}
	\affiliation{School of Chemistry, Faculty of Exact Sciences $\&$ The Center for Physics and Chemistry of Living Systems, Tel Aviv University, Tel Aviv 6997801, Israel}

	%\affiliation{Division of Dynamics, Faculty of Mechanical Engineering, Lodz University of Technology, 90-924 Lodz, Poland}
	\author{Dibakar Ghosh}
	%\email{chittaranjanhens@gmail.com}
	\affiliation{Physics and Applied Mathematics Unit, Indian Statistical Institute, 203 B. T. Road, Kolkata 700108, India}		
	\author{Syamal K. Dana}	
	\affiliation{Centre for Mathematical Biology and Ecology, Department of Mathematics, Jadavpur University, Kolkata 700032, India}
	\affiliation{Division of Dynamics, Faculty of Mechanical Engineering, Lodz University of Technology, 90-924 Lodz, Poland}
	%\affiliation{Department of Mathematics, Jadavpur University, Kolkata 700032, India}
	\author{Chittaranjan Hens}
	\email{chittaranjanhens@gmail.com}
	\affiliation{Physics and Applied Mathematics Unit, Indian Statistical Institute, 203 B. T. Road, Kolkata 700108, India}
	\date{\today}

	\begin{abstract}
How long does a trajectory take to reach a stable equilibrium point in the basin of attraction of a dynamical system? This is  a question of quite general interest, and has stimulated a lot of activities in dynamical and stochastic systems where the metric of this estimation is often known as the transient or first passage time. In nonlinear systems, one often experiences long transients due to their underlying dynamics. We apply resetting or restart, an emerging concept in statistical physics and stochastic process, to mitigate the detrimental effects of prolonged transients in deterministic dynamical systems. We show that stopping an ongoing process at intermittent time only to restart all over from a spatial control line, can dramatically expedite its completion, resulting in a huge decrease in mean transient time. Moreover, our study unfolds a net reduction in fluctuations around the mean. Our claim is established with detailed numerical studies on the Stuart-Landau limit cycle oscillator and chaotic Lorenz system under different resetting strategies. Our analysis opens up a door to control the mean and fluctuations in transient time by unifying the original dynamics with an external stochastic or periodic timer, and poses open questions on the optimal way to harness transients in dynamical systems.
	\end{abstract}
	\maketitle

Transient time  is unequivocally an important attribute of dynamical systems. In simple words, transient time (TT) quantifies the time it takes for a trajectory to reach from any point P to another point Q, specifically from an initial state to an attractor i.e, stable oscillation or an equilibrium point.  In recent times, statistics of TT has been extensively studied in  complex dynamical systems \cite{yorkeprl1986, lai2011transient, yorke1979,altmann2013leaking,  parlitzPRL2017, parlitzPRL2018}, climate models \cite{lenton2011early,scheffer2009early}, ecology \cite{hastings2018transient, morozov2019long, gosztolai2019collective,martin2020importance}, signal propagation in networks \cite{hens2019spatiotemporal,tarnowski2020universal} and extreme events like catastrophes or species extinction \cite{hastings2004transients,hastings2010timescales,majumdar2020extreme}. TT has also been a key ingredient to understand critical transitions from one stable ecosystem state to another often known as tipping \cite{vanselow2019very} or regime shift \cite{scheffer2001catastrophic}. In ecology, faster convergence to stable solutions under external perturbations is known to be of severe importance to sustain resilience \cite{arnoldi2018ecosystems,gao2016universal}. Similarly, one can ask whether it is possible to operate a power grid network \cite{motter2013spontaneous} with a faster realization of synchrony to avoid a failure. Thus, the intriguing questions are how to tailor \textit{generic strategies} to understand optimization and control of transient time in natural and engineered systems.

\par Transient time is also a subject of immense interest in statistical physics and stochastic process. Therein, it is often known as the first passage time (FPT) which measures the completion time of a process (see \cite{redner2001guide,bray2013persistence,metzler2014first,benichou2011intermittent} for extensive reviews). Despite many years of rigorous studies, efforts are still being made in search of finding new protocols to make the FPT processes more efficient \cite{PhysRevLett.124.090603,benichou2010geometry,condamin2007first,levernier2019survival,guerin2016mean}. Recently, it has been observed that completion of a FPT process can be expedited by resetting it intermittently and starting afresh \cite{evans2020stochastic,evans2011diffusion,evans2011diffusion-opt,reuveni2016optimal,pal2017first,pal2019first,belan2018restart,pal2016diffusion,reuveni2014role,luby1993optimal,montanari2002optimizing,kusmierz2014first,falcon2017localization,chechkin2018random,lapeyre2017reaction,pal2019landau,boyer2014random,bhat2016stochastic}. 
%A flurry of followup results have further established this identification. 
This problem is known as first passage under restart or resetting and has led to a myriad of interesting phenomena with an overarching stream of applications in non-equilibrium systems \cite{evans2020stochastic}, biological and chemical processes \cite{reuveni2014role,lapeyre2017reaction}, randomized search algorithms in computer science \cite{luby1993optimal,montanari2002optimizing}, search and foraging theory \cite{falcon2017localization,chechkin2018random,boyer2014random,bhat2016stochastic}. The pinnacle of these studies is perhaps the expedition of the mean FPT by choosing a careful restart mechanism.

 Despite a wide array of studies made in noisy systems, a little knowledge exists, in literature, on the impact of resetting strategies in deterministic dynamical systems.  For instance, resetting can be understood as restoration of an apex predator or other species population in the hierarchical levels of a food chain for biodiversity conservation \cite{ritchie2009} or a catastrophe \cite{dharmaraja2015continuous}. Naturally, the question arises whether resetting 
 can now be used as a \textit{control strategy} for TT in deterministic dynamics where the target is a stable steady state, a limit cycle or a chaotic orbit. Furthermore, it is not apparent how to implement the resetting mechanism since the intrinsic dynamics is deterministic and thus, restarting the system from the same initial condition can not improve the transient time. To address these challenges, in this article, we numerically study TT in the presence of resetting. We seek for efficient protocols based on resetting to mitigate the effects of long  transient time in dynamical systems having stable equilibrium points and furthermore strive to  make them optimal. In particular, our results are illustrated with two canonical models of deterministic systems namely a Stuart-Landau oscillator (denoted by $\M_1$) and the Lorenz system (denoted by $\M_2$). The central finding of our study reveals that resetting on a spatial \textit{control line}  which is constructed arbitrarily through the stable equilibrium point(s) in the basin of attraction dramatically \textit{reduces the mean and fluctuations} in transient time and thus outperforms the completion.

\begin{figure}[t]
	%	\centerline{
	\includegraphics[scale=0.25]{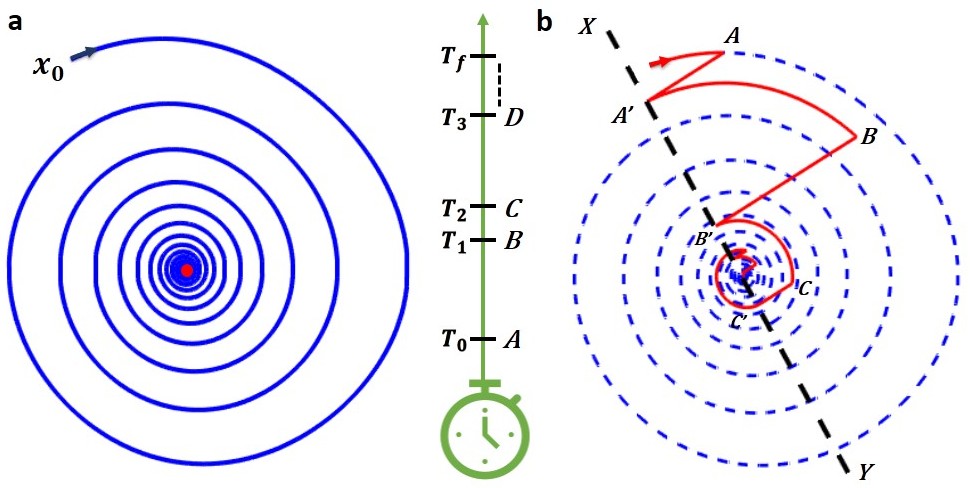}%
	\caption{Illustrative description of resetting strategy. (a) Trajectory of a deterministic system that starts from the initial state $\textbf{x}_0$ and reaches an equilibrium point (red dot). (b) Trajectory under resetting is projected momentarily ($A\to A'$, $B\to B'$ and so on) on the control line  $XY$ (black dashed line) chosen from the basin $\mathcal{B_A}$ and passing through the equilibrium point. Resetting events occur at random times $T_0$, $T_1$, $T_2, \cdots$, $T_f$ (as shown by the clock), where the time intervals $T_i-T_{i-1}$ are taken from a distribution $f_{R}( t)$. The trajectory with resets (red line) is superimposed on the unhindered trajectory (dashed blue line).}
	\label{fig1_Illus} 
\end{figure}

\textit{Transient time.---} Consider an autonomous system spanned in a basin $\mathcal{B_{A}}$ and described by 
\begin{eqnarray}
\label{eq.1}	
\dot{\textbf{x}}  = {F}(\textbf{x},\boldsymbol{\mu}),
%\end{array}
\end{eqnarray}
where $\textbf{x}$ is the state variable, $F$ 
is a smooth vector field with dimension $n$ and
 $\boldsymbol{\mu}$ is the system parameter. If the system has a monostable point attractor i.e., a stable equilibrium point $\A$ (say, the target point), then $TT(\textbf{x}_0)$ is the time required, in the absence of resetting, to reach the stable attractor of the system from randomly chosen initial points $\textbf{x}_0\in \mathcal{B_{A}}$. Following \cite{lucarini2016extremes,lundstrom2018find,klinshov2018interval,kittel2017timing}, the metric for $TT$
is defined as
 \begin{eqnarray}
TT(\textbf{x}_0)=\inf  \{t:\|\phi^{TT}(\textbf{x}_0)- \A\|<\epsilon,  \},
\label{TT_1}
\end{eqnarray} 
where $\| . \|$ denotes the Euclidean distance and we assume that the system evolves through a time evolution operator $\phi$ and reaches to $\phi^{TT} (\textbf{x}_0)$ at a time $t>0$. Further, we set $\epsilon$ with a pre-defined threshold
which is chosen arbitrarily small so as to characterize an approximate proximity of the numerical trajectory to the asymptotically stable equilibrium point $\A$. The set of transient time over the initial conditions in $\mathcal{B_{A}}$ is then simply given by $\{TT(\textbf{x}_0), ~ \forall \textbf{x}_0 \in \mathcal{B_{A}} \}$ (Sec. I in \cite{SM}).

\begin{figure*}
\includegraphics[scale=0.28]{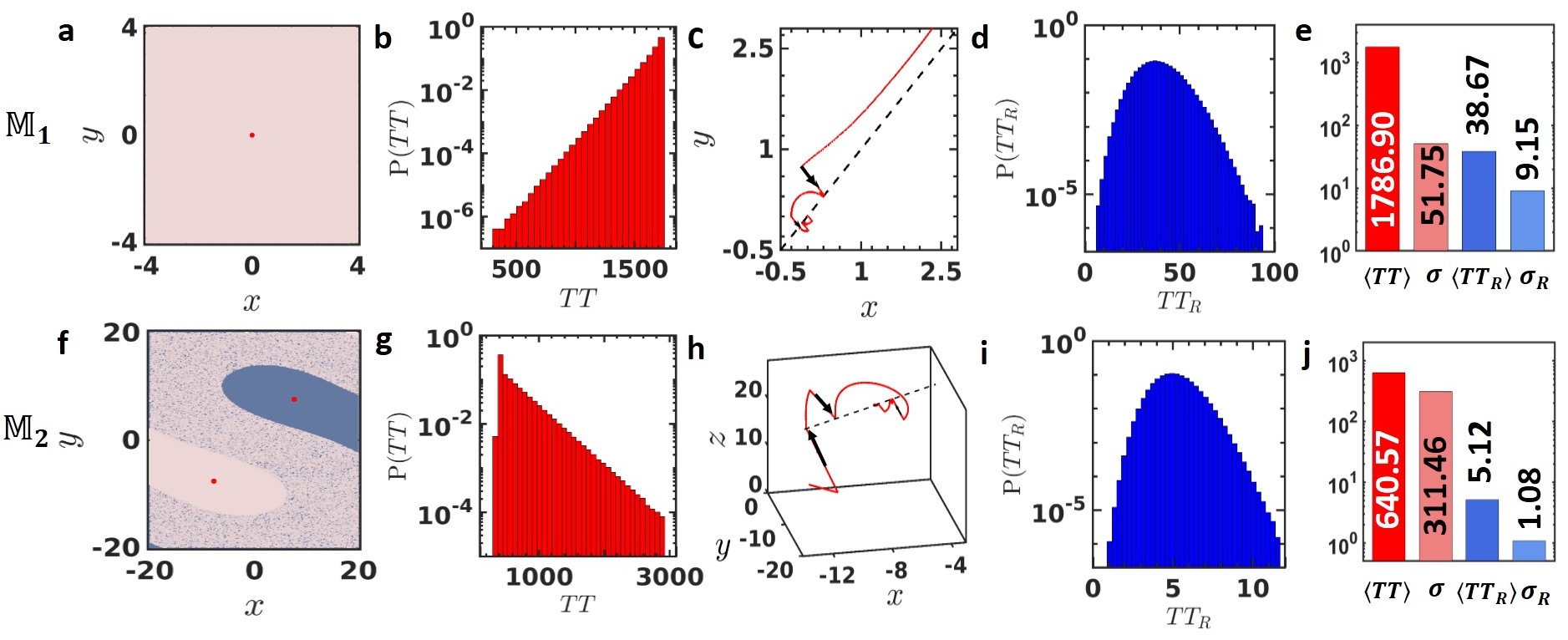}
	\caption{Basin of attraction:  $\mathcal{B_{A}}$ 
		with $\mathcal{A}$ (red dot) for the system (a) $\M_1$ and (f)	$\M_2$.
	  Transient time density without resetting: $P(TT)$ for  (b) 
		$\M_1$ and (g) $\M_2$.
		Phase space trajectories under resetting: Trajectories of the system (c)	$\M_1$, starting from the co-ordinates $(2.5,3.0)$
		and (h) $\M_2$, starting from the co-ordinates $(-11.74,-4.412,4.086)$ are depicted by red lines. The black arrows indicate the resets or normal projections to the control line (dashed black line passing through $\mathcal{A}$). Resetting occurs at $\langle R \rangle=1$ and $\langle R \rangle=0.1$, respectively, for $\M_1$ and $\M_2$.
		  Transient time density with resetting: $P(TT_R)$  in the presence of exponential resetting for the systems, (d) 
		$\M_1$ and (i) $\M_2$, respectively, with aforementioned resetting rates.
		Comparison of mean and fluctuations in transient time without and with resetting: Reduction in $\langle TT_R \rangle$ and $\sigma_R$ is observed in the bar plots (with a comparison between their corresponding values)
		 for (e) $\M_1$ 
		and (j) $\M_2$ respectively.}
	\label{fig1} 
\end{figure*}

\textit{Resetting protocol \& control line.---} 
Canonical restart mechanisms, in statistical physics, usually set the configuration of a system to its initial state after a random time which is drawn from a distribution given by $f_R(t)$ \cite{evans2020stochastic,gupta2014fluctuating,pal2015diffusion,mendez2016characterization,VV2019landau,eule2016non,gupta2020work}. Herein, resetting the process to the initial condition is an impediment due to the strong determinism encoded in the underlying dynamics. To circumvent this issue, we reset or project the dynamics along a line, which we define as a \textit{control line} in $\mathcal{B_A}$ but passing through the equilibrium point $\A$. To illustrate the concept, we refer to \fref{fig1_Illus}a, where we have considered a trajectory of the uninterrupted process that starts from the initial condition $\textbf{x}_0$. $TT(\textbf{x}_0)$ denotes the time required by the trajectory starting from $\textbf{x}_0$ to reach the fixed point (red dot) within a precision of $\epsilon$. The control line $XY$ (black dashed line, Fig.\ \ref{fig1_Illus}b) is constructed at a random angle $\theta \in (0,2\pi)$ but passing through $\mathcal{A}$ (red dot). We stop the dynamics e.g., at time $T_0$ drawn from $f_R(t)$ and reset the current position (say, $A$) to a point (say, $A'$) on the control line by projecting it normally. Subsequently, the dynamics starts from the point $A'$. The next time interval $T_1-T_0$ is again drawn from the density $f_R(t)$ and the procedure is repeated. The resulting trajectory  after several resets (i.e., with projections on the control line) at coordinates 
$A, B, C$ and $D$ is shown by the solid line (red line). The process ends when the condition $\|\phi^{TT}(\textbf{x}_0)- \A\|<\epsilon$ is satisfied for \textit{the first time} and we denote this net transient time as $TT_R$. 
Against this backdrop, we study statistics of $TT_R$ with different choices of $f_R(t)$ for systems,  $\M_1$ and $\M_2$, which we introduce now in brief.

\textit{Stuart-Landau $(SL)$ oscillator} ($\M_1$).---
SL oscillators are abundantly used to understand many fundamental phenomena such as transition to synchrony and pattern formation \cite{kuramoto2003chemical,pikovsky2003synchronization}.
The governing equation for such an oscillator reads $\dot{Z}  = (a + i\Omega-|Z|^2)Z,$
where $Z=x+iy$ is the complex variable and $\Omega$ is the natural frequency of oscillation. Here, $a$ is an internal control parameter that determines the state of the system (oscillatory or a steady state). Initial conditions are chosen uniformly from $\mathcal{B_A}$ in \fref{fig1}a. Following a linear stability analysis (Sec. II in \cite{SM}), it is shown that the system exhibits a stable spiral approaching an equilibrium point $\mathcal{A}: (0,0)$ for $a < 0$ and stable limit cycle for
 $a \geq 0$. We set  $a=-0.01$ and $\Omega=1$ so that the system, after a transient time, attains to $(0,0)$ shown by the red dot in \fref{fig1}a.
 
  \textit{Lorenz system} ($\M_2$).--- Lorenz system is a benchmark model of chaotic systems \cite{lorenz1963deterministic,ott2002chaos,strogatz2001nonlinear}.
Here, the phase space equations are
$\dot{x}  = \sigma(y-x), \dot{y}  = \rho x-y-xz, \dot{z}  = -\beta z+xy,$
where $\sigma$ and $\rho$ are the Prandtl and Rayleigh numbers, respectively, while $\beta>0$ is the aspect ratio. The system has two symmetric stable equilibrium points $\left(\pm \sqrt{\beta(\rho-1)},\pm \sqrt{\beta(\rho-1)},\rho-1\right)$ only if $1<\rho <\dfrac{\sigma(\sigma+\beta+3)}{\sigma-\beta-1}$.
For fixed parameters $\sigma=10$ and $\beta=\frac{8}{3}$, the system exhibits transient chaos in the range of $\rho \in (13.926,24.06)$ \cite{SM}. Considering 
 $\rho=23$,  we obtain two stable fixed points $\mathcal{A}:\{ P_{1,2}=(\pm a,\pm a,b)   \}$, where $a=7.65942,~ b=22.0$. We observe a riddled basin, with two disjoint basins  of attraction for two emerging scrolls in the dynamics, surrounding two separate equilibrium points (red dots) (Fig.\ \ref{fig1}f).

\textit{Statistics of transient time without resetting.---} To elucidate the effects of resetting, it is important to first study the transient time statistics of the underlying systems. To this end, 
we simulate $\M_1$ and $\M_2$ using the $4^{th}$ order Runge-Kutta method while starting from their individual basin of attraction (\cite{SM}). $\M_1$  is a monostable system and has stable spiral trajectory while $\M_2$ is a 3-dimensional bistable system in which two stable fixed points appear together with two separated and intermingled basins. The system either converges to a single fixed point (for $\M_1$) or fixed points (for $\M_2$) followed by a damped oscillation or a transient chaotic phase for the chosen parameters. Integrating the systems  from $5 \times 10^6$ initial conditions, we have tracked the entire set of reaching time to the  vicinity  of stable equilibrium points following the condition (given by Eq. \ref{TT_1}) with $\epsilon=10^{-9}$ set for both the models.  To capture the appropriate statistics of $TT$, we have scanned the entire basin with a finite resolution, however, discarding the initial conditions which set off from a distance smaller than $10^{-5}$ from the targeted fixed point. The resulting density functions are shown in \fref{fig1}b ($\M_{1}$) and \fref{fig1}g  ($\M_{2}$). 
%We have plotted the density of $TT$ \st{(averaged over these initial conditions)} in \fref{fig1}b (for $\M_1$) and \fref{fig1}g (for $\M_2$).
 We observe from \fref{fig1}b that $P(TT)$ is supported from above. This is because, in $\M_1$, $TT$ increases exponentially as a function of the Euclidean distance between the initial and targeted state before it saturates to a threshold point which in turn corresponds to the upper bound (\cite{SM}). Note that such a relationship is not pertinent to model $\M_2$. However, there the transient time is exponentially distributed (\fref{fig1}g) which is a characteristic feature of chaotic systems \cite{yorkeprl1986, yorke1979}.

\textit{Transient time under resetting.---} To employ resetting on the underlying dynamics, we first  choose the resetting time density to be exponential so that $f_R(t)=\langle R \rangle^{-1} e^{-t/\langle R \rangle}$ which essentially means that resetting occurs at a rate $1/\langle R \rangle$ \cite{evans2020stochastic}. As outlined before, we first construct the control line in each case and set them fixed for the entire simulation.
For $\M_1$, the control line is chosen diagonally along the basin and passes through $\A: (0,0)$ (black dashed line in \fref{fig1}c). Figure \ref{fig1}c shows a representative trajectory (red line) with multiple  attempts of resetting (black arrows) at a rate $\langle R \rangle^{-1}=1.0$. The resulting distribution of $TT_R$ (Fig.\ \ref{fig1}d) immediately reveals two key observations: reduction in both the mean transient time $\langle TT_R \rangle$ and fluctuations $\sigma_R \equiv \sqrt{\langle TT_R^2 \rangle-\langle TT_R \rangle^2}$ around the mean.
% i.e., $\langle TT_R \rangle <\langle TT \rangle$. 
Here, for the current choice of parameters, we noted a dramatic speed up of $\sim 46$ and $\sim 6$ times for the mean and fluctuations, respectively (\fref{fig1}e).

 \begin{figure}[t]
	%	\centerline{
%	\includegraphics[scale=0.2]{figure_2v2.jpg}
\includegraphics[height=3.25cm,width=8.75cm]{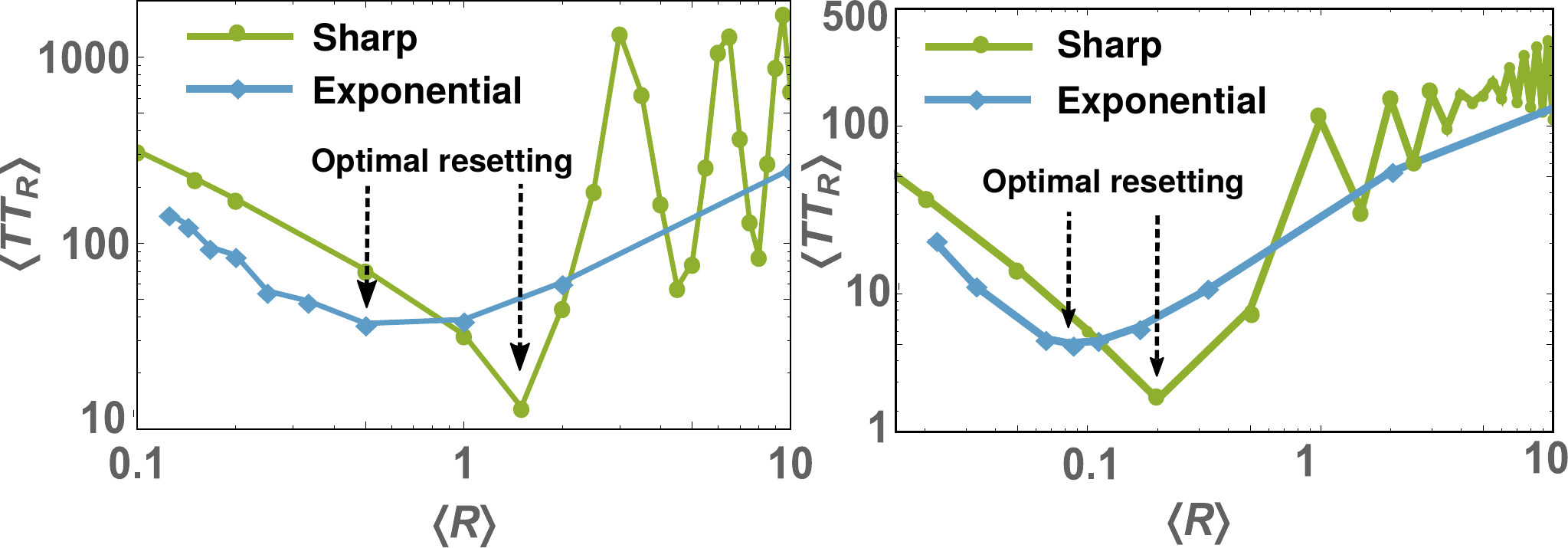}
	\caption{Plot of $\langle TT_R \rangle$ as a function of $\langle R \rangle$ for $\M_1$ (left panel) and $\M_2$ (right panel). Resetting times are chosen from exponential (diamond markers) and sharp distribution (circle markers). Sharp resetting reduces $\langle TT_R \rangle$ more efficiently than the exponential at the optimal time $\langle R^* \rangle$.} 
	\label{fig2} 
\end{figure} 
 
 A similar picture is delineated for model $\M_{2}$ in Fig.\ \ref{fig1}h-j. In this case, there is some  freedom in the choice of the control line since we have two equilibrium points (which are also the targets) $P_1$, and $P_2$. 
The control line can be drawn through either of the equilibrium points or connecting both. 
We choose the latter case and the resetting procedure is conducted identically at a rate $\langle R \rangle^{-1}=10$ (for the details of the former choice, Sec.\ V in \cite{SM}). Collecting the data statistics of $TT_R$, we plot a histogram in \fref{fig1}i, which estimates that resetting over-performs the mean
by $\sim 125$ folds. Alike $\M_1$, we find a significant reduction ($\sim 288$ times) in the fluctuations for $\M_2$ (\fref{fig1}j).

 To show that indeed this behavior is generic, we now adapt a different strategy where resetting takes place always after a fixed time $\langle R \rangle$ so that $f_R(t)=\delta(t-\langle R \rangle)$. This is often known as the sharp resetting which was proven to be the most time-efficient protocol in stochastic systems \cite{pal2016diffusion,pal2017first,chechkin2018random}. Again, the highlighting features here are the decrements in mean and fluctuations in transient time (see \fref{fig2}). Manifesting the control line  protocol, we find that sharp restart reduces the mean and fluctuation by $\sim55$ and $\sim152$ folds for $\M_1$ when $\langle R \rangle=1$. Similarly, for $\M_2$, we observe a speed up of $\sim106$ for the mean and $\sim438$ folds for the fluctuations when performing at a rate $\langle R \rangle=0.1$ ( \cite{SM}).

 To delve deeper,  we now scan $\langle TT_R \rangle$ as a function of $\langle R \rangle$ in \fref{fig2} for both the resetting schemes. When $\langle R \rangle$ is small, the system  resets too frequently so that the trajectory is effectively confined near the control line and the transient time is achieved by these short excursions. On the other hand, when $\langle R \rangle$ is large, the waiting time between resetting events increases. In other words, there is hardly any resetting event and the completion is achieved typically by the original dynamics. For sharp resetting (circle marked green lines), markedly distinct oscillatory behavior emerges when $\langle R \rangle$ often becomes the integer multiple or half integer multiple of the intrinsic time period of $\M_1$ (Sec.\ IV in \cite{SM}). This happens since sharp resetting is a periodic process and is always conducted after a fixed time $\langle R \rangle$. On the other hand, for $\M_2$, we do not observe any systematic pattern due to its aperiodic nature. For exponential resetting, variation of $\langle TT_R \rangle$ as a function of $\langle R \rangle$ are shown by the diamond marked blue lines in the same figure where the qualitative features are found to be similar. However, oscillations are not present here since $f_R(t)$ is a continuous distribution and thus the waiting time between resetting events are not multiples of the underlying time period.

 In the intermediate regime of $\langle R \rangle$, in both the cases, the trajectory explores its intrinsic dynamics between consecutive resetting events. The combined effect essentially leads to a drastic decrease in $\langle TT_R \rangle$ (see \fref{fig2}). Quite interestingly, we see emergence of an optimal resetting rate $\langle R^{*} \rangle$ such that $\frac{d\langle TT_R \rangle}{d\langle R \rangle} |_{\langle R^{*} \rangle}=0$. In our set up, we find that for $\M_1~ (\M_2)$, the optimal transient times $\langle TT_R^{*}\rangle$ for the exponential and sharp resetting are $\approx 36.86 ~(5.06)$ and $\approx 13.5 ~(2.35)$ respectively. The above analysis clearly indicates that sharp resetting could work more efficiently to reduce transient time than the exponential resetting at the optimal condition.

\textit{Discussions and future outlook.---}
In this paper, we showcase a first study on the application of resetting in deterministic dynamical systems having prolonged transient time. We show that systematic controlled resetting strategies, which mix and match external stochastic and periodic timers with internal spatial properties, have an  ability to facilitate the completion of a process, by reducing mean and fluctuations, which otherwise would hinder. With the aid of numerical simulations, we investigate two paradigmatic non-linear systems under Poisson or exponential and sharp resetting. Noteworthy in this regard is the \textit{dominance of sharp resetting over exponential resetting} at the optimality. While this observation is quite intriguing, future studies to formally establish this result in dynamical systems look like a serious challenge.

To conceptualize resetting in our systems, we have introduced the notion of a \textit{control line} to which the system is projected after each resetting. We have shown that the method of control line performs proficiently for both the models and thus is quite robust. For a homogeneous basin ($\M_1$), the reduction in transient time remains fully invariant on the choice of control line (Sec.\ V in \cite{SM}). However, for $\M_2$, the transient time depends clearly on the choice of the control line which here can be of three kinds passing through $P_1$ or $P_2$ (or both). For the first two cases, the system reaches to their respective equilibrium points while for the third case the probability to converge to any of these equilibrium points is equally shared. It is important to point out that the models chosen here show behavioral shift (steady state to oscillation, periodic or chaotic) when we change the system parameters to a critical value. Remarkably, even near the onset of critical transitions, we find that resetting remains beneficial for a range of parameters (Sec.\ VI in \cite{SM}).

Concluding, we stress that we have shown extensively that persistent resetting can reverse the deleterious effects of long transient time in autonomous systems. Notably, in this first case study, we have assumed resetting process to be instantaneous in order to keep congruence with the original idea of resetting. However, to adapt realistic scenarios, future studies need to be carried out to explore the effects of a time overhead or delay due to resetting \cite{pal2019invariants}. Nonetheless, we believe that the qualitative key features observed here should remain invariant. Thus, indeed, resetting
can operate as a powerful assay to regulate transient time in complex systems.

\textit{Acknowledgments.---}
The authors would like to thank Sarbendu Rakshit for interesting discussions and notable comments. A. P. gratefully
acknowledges support from the Raymond and Beverly
Sackler Post-Doctoral Scholarship at Tel-Aviv University. C.H. is supported by
DST-INSPIRE Faculty Grant No. IFA17-PH193.
\onecolumngrid
\pagebreak

\begin{center}
	\textbf{\Large Supplemental Material:  ``Mitigating long transient time in deterministic systems by resetting}
\end{center}

\tableofcontents

\newpage 
\renewcommand{\arraystretch}{1.5}
\allowdisplaybreaks[1]
\newcommand{\Sref}[1]{Section~\ref{#1}}%
\newcommand{\sgn}[1]{\mathrm{sgn}({#1})}%
\newcommand{\erfc}{\mathrm{erfc}}%
\newcommand{\erf}{\mathrm{erf}}%

\newcommand{\lk}[1]{\textcolor{olive}{#1}}
\def\hM{{\bm{M}}}
\def\hK{{\bm{K}}}
\def\hg{{\boldsymbol{\gamma}}}
\def\hG{{\boldsymbol{\Gamma}}}
\def\bS{{\boldsymbol{\Sigma}}}
\def\bG{{\bm{G}}}
\def\bg{{\bm{g}}}
\def\bea{\begin{eqnarray}}
	\def\eea{\end{eqnarray}}
\def\etal{et al.}
\def\nn{\nonumber}
\def\om{\omega}
\def\f{\frac}
\def\bn{{\bm{n}}}
\def\be{{\bm{\hat{e}}}}
\def\bell{\bm \ell}
\def\br{\bm r}
\def\bZ{\bm Z}
\def\p{\partial}

\section {Notation and definition of  Transient Time}\label{Notation} 
A deterministic dynamical system can be captured by an ordinary differential equation of the following form
\begin{eqnarray}
	%\begin{array}{l}
	\label{eq.s11}	
	\dot{\textbf{x}}  = {F}(\textbf{x},\bm{\mu}),
	%\end{array}
\end{eqnarray}
where  $F$ 
represents the vector field, $\textbf{x}\in\R^{n}$ and $\bm{\mu}$ is the parameter. 
%and specifies/maps the evolution of the system dynamics within the $n$- dimensional state space. 
Let $\A~(\subseteq {\R}^{n})$ be an attractor of the Eq.\ (\ref{eq.s11}) and corresponding basin of attraction is denoted by 
$\mathcal{B_{A}}~(\subseteq {\R}^{n})$. Let $\textbf{x}_0=(x_{10},x_{20},...,x_{n0})^\mathcal{T}$ ($\mathcal{T}$ denotes the transpose of a matrix) ~%(\in \mathcal{B_{A}})$ 
be an initial condition at $t=t_0$ from which the system evolves through a time evolution map $\phi$ and reaches to $\phi_{t_0}^t (\textbf{x}_0)$ at  time $t>0$. 
%Here 
%$\phi^t$ is the time-evolution map.  
If $\A$ is an asymptotically stable equilibrium point,  we can write $\phi_{t_0}^{t \to \infty} (\textbf{x}_0) \rightarrow \A$. 
%as 
%$t \to \infty$. 
For practical purpose, we assume that the system reaches to the close vicinity of the stable attractor $\A$ in a finite time, say $TT(\textbf{x}_0)$ for initial state $\textbf{x}_0$. We call this finite time $TT(\textbf{x}_0)$ as {\it transient time}, which is formally defined as follows
%\noindent
\begin{eqnarray}
	\begin{array}{l}\label{tt}	
		TT(\textbf{x}_{0})=\inf \{t: \| \phi^{TT}(\textbf{x}_{0})- \A\|<\epsilon\},
	\end{array}
\end{eqnarray}
where $\epsilon$ is a small positive number and $\| . \|$ denotes the Euclidean distance. One can define this metric $D$ as $D(\textbf{x},\textbf{y})=\sqrt{\sum_{i=1}^{n}(x_i-y_i)^2}$, where $\textbf{x}=(x_1,x_2,...,x_n)\in \R^n$ and $\textbf{y}=(y_1,y_2,...,y_n)\in \R^n$.  We set $\epsilon=10^{-9}$ for our simulations.
%\begin{eqnarray}
%	\begin{array}{l}	
%TT(\textbf{x}_{0})=\{t: \| \phi^t(\textbf{x}_{\infty})- \A\|<\epsilon, ~ \textbf{x}_{\infty} \in \mathcal{B_{A}} \},
%\end{array}
%\end{eqnarray}
%$TT(\textbf{x}_0)$ 
%is required time for a system from any initial position 
%$\textbf{x}_0$ 
%to reach to the vicinity of the stable attractor 
% $\A$.
%Theoretically a system possesses infinite time to reach its asymptotically stable attractor. But to deal with real life practical problem or numerical simulation, we must truncate the reaching time at a finite value under which time system can arrive at $\ A$, by using a threshold (say, $\epsilon$) over the distance between  $\phi^t (\textbf{x}_0)$ and $\A$.
% . So, if time  , final state or target, then for a pre-defined $\epsilon (>0)$ $TT(\textbf{x}_0)$ is defined as 
%{\color{blue} Then, with probability $p$,  we reset the evolution map $\phi$ onto an arbitrary straight line in the phase space passing through the stable equilibrium point. The probability of resetting is $p=r h$, where $r$ is the resetting rate and $h$ is the integration time step.  Large $r$ indicates very fast resetting of the flow on the line, whereas small $r$ implies that the resetting occurs very slowly, as the projection on the line has a very low probability.} 
Now, the set of transient time over the entire basin
$\mathcal{B_{A}}$ 
can be constructed as
$	 \{TT(\textbf{x}_0), ~ \text{for all}~ \textbf{x}_0 \in \mathcal{B_{A}} \},$ i.e. $\textbf{x}_0$ is all accessible initial conditions  in the basin $\mathcal{B_{A}}$.\\
%Now, the calculated transient time after resetting processes is denoted by $TT_{R}$ and defined as   
%\begin{equation*}
%\begin{array}{l}	
%TT_R(\textbf{x}_0,r)=\inf \{t: \| \phi^t(\textbf{x}_0)- \mathcal{A}\|<\epsilon, ~ \textbf{x}_0 \in \mathcal{B_{A}}, r \in (0, r_{max}] \},
%\end{array}
%\end{equation*}
%{\color{blue} where $r_{max}$ is the maximum resetting time when the probability of resetting $p=1.0.$}
%This calculated finite value of transient time depends on choosing of $\epsilon$. In this article, we choose $\epsilon=1.0 \times 10^{-9}$ to calculate the transient time.         
\section{Linear stability analysis: $\M_1$ and $\M_2$}\label{Stability}
In this section, we present a linear stability analysis for the two paradigmatic models used in the main text namely the Stuart-Landau system $(\M_1)$ and the Lorenz system $(\M_2)$.
\subsection {Eigenvalue analysis of $\M_1$}
\label{Eigen_M_1}
\noindent
Stuart Landau model $(\M_1)$ is described by the following governing equation of motion \cite{ott2002chaos,strogatz2001nonlinear}
\begin{eqnarray}
	\begin{array}{l}\label{eq.s1}	
		\dot{Z}  = (a + i\Omega-|Z|^2)Z,
	\end{array}
\end{eqnarray}
where $Z=x+iy$ is the complex variable;  $a$  and $\Omega$ are the intrinsic parameters of the system. The system has one equilibrium point at $(0,0)$. Now, the Jacobian matrix $J$ of the system $\M_1$ at the  equilibrium point $(0,0)$ is given by
\begin{eqnarray*}
	J(0,0)=
	\left[ {\begin{array}{ccc}
			a & -\Omega  \\
			\Omega & a  \\
	\end{array} } \right].
\end{eqnarray*}
The   characteristic roots of the above Jacobian are 
$\lambda_{\pm}=a\pm\Omega i=-0.01\pm i$, where 
$a=-0.01$ and $\Omega=1$. Therefore, the trivial equilibrium point
$(0,0)$ is a stable spiral. 
% We know the real part of eigen value decides the decaying nature of oscillations {\it i.e.} how fast a system reaches to the stable attractor.}| 
Here, the system parameter $a$ determines decay rate. On the other hand, the imaginary part of the eigenvalue  $\Omega$ determines the intrinsic frequency of this decaying oscillation.  Thus, the  time period of oscillatory behavior during the transient phase, for our current choice of parameters, is given by
\begin{eqnarray}
	T(\M_1)\sim\frac{2\pi}{\Omega}\approx6.28318.
	\label{T_M1}
\end{eqnarray}
We note that the system experiences a critical transition (from stable spiral to a stable limit cycle) at $a_c\equiv a=0.0$.

\subsection {Eigenvalue analysis of $\M_2$}
\label{Eig_M2}
\noindent
The governing equation of motion for the Lorenz system ($\M_2$)  is given by \cite{ott2002chaos,strogatz2001nonlinear}
\begin{equation}
	\begin{array}{l}\label{eq.s2}	
		\dot{x}  = \sigma(y-x),\\
		\dot{y}  = \rho x-y-xz,\\
		\dot{z}  = -\beta z+xy,\\
	\end{array}
\end{equation}
where the system parameters are $\sigma, \rho,$ and $\beta$ $(>0)$. It is easy to see that the system has  a trivial equilibrium point $P_0: (0, 0, 0)$ which is stable for $\rho<1.$ For $\rho>1$, two non-trivial equilibrium points emerge which are given by $P_{1}:(\sqrt{\beta(\rho-1)},\sqrt{\beta(\rho-1)},\rho-1)$ and $P_{2}:(-\sqrt{\beta(\rho-1)},-\sqrt{\beta(\rho-1)},\rho-1)$.
Now we proceed to calculate the Jacobian matrix $J$ of the system $\M_2$ at the equilibrium point  $P_{1}$. This gives
\begin{eqnarray}\label{jacobian}
	J(\sqrt{\beta(\rho-1)},\sqrt{\beta(\rho-1)},\rho-1)=
	\left[ {\begin{array}{ccc}
			-\sigma & \sigma  & 0 \\
			1 & -1  & -\sqrt{\beta(\rho-1)} \\
			\sqrt{\beta(\rho-1)} & \sqrt{\beta(\rho-1)}  & -\beta \\
	\end{array} } \right].
\end{eqnarray}
One can now immediately write the characteristic equation coming from the Jacobian above, and this reads
\begin{equation}\label{jacobian1}
	\lambda^3+(\beta+\sigma+1)\lambda^2+\beta(\rho+\sigma)\lambda+2\beta\sigma(\rho-1)=0.
\end{equation}
For fixed parameters e.g., $\sigma=10,\beta=\dfrac{8}{3}$ and $\rho=23$, the  characteristic equation \eref{jacobian1} becomes
\begin{equation}\label{jacobian2}
	\lambda^3+\dfrac{41}{3}\lambda^2+88\lambda+\dfrac{3520}{3}=0.
\end{equation}
The roots of  the above equation are simply given by $\lambda=-13.5588; -0.054\pm 9.3024 i$.
Therefore linear stability analysis at the vicinity of $P_1$ determines that it is a stable spiral. %We get characteristic equation same as Eq.\ (\ref{jacobian1}) of the Jacobian $J(-\sqrt{\beta(\rho-1)}, -\sqrt{\beta(\rho-1)},\rho-1)$ for equilibrium point $FP_2$. So, 
In the same way, one can also show that $P_2$ is a stable spiral.
Note that, the system has a transient chaos phase in a range of $\rho\in(13.926,24.06)$ for $\sigma=10$ and $\beta=\dfrac{8}{3}$ \cite{yorkeprl1986}. 
Increasing  $\rho$ towards the critical transition point ($\rho_c=24.06$), the  duration of chaotic transient phase follows a power law \cite{yorkeprl1986,lai2011transient,yorke1979}. At $\rho=\rho_c$, the critical transition occurs and the transient chaos becomes a chaotic attractor. %We have  discussed the role of the parameter  $\rho$ on transient time under resetting in Sec.\ \ref{parameter}.

\section{Distance and transient time density without resetting}\label{Distance}
In this section, we discuss in details the quantitative features of the transient time density $P(TT)$ in the absence of resetting. To obtain the histogram for each model, we have scanned $5\times10^6$ initial conditions from the basin of attraction $\mathcal{B}_{\A}$.

\subsection{Transient time for $\M_1$}
In the case of system $\M_1$, we choose our basin span to be 
$[-6,6]\times[-6,6]$, and collect the transient time. The resulting density is plotted in Fig.\ 2b in the main text. From the figure, it becomes evident that the density is supported from above. Moreover, we observe that the
probability to get larger values of $TT$ is higher than the smaller values of $TT$. To gain deeper insights, we have investigated the relation between the transient time (of the trajectories taken from initial points in basin to stable equilibrium point) and the Euclidean distance (between initial and target states). 
The Euclidean distance, metric $D$, is described as
\begin{eqnarray}
	D(\textbf{x},\textbf{y})=\sqrt{\sum_{i=1}^{n}(x_i-y_i)^2}~,   \end{eqnarray}
where $\textbf{x}=(x_1,x_2,...,x_n)\in \R^n$ and $\textbf{y}=(y_1,y_2,...,y_n)\in \R^n$. We collect all $D$ and $TT$ for both models and plot them  in Fig.\ \ref{figs1}a. For $\M_1$, the transient time increases exponentially as we increase the Euclidean distance ($TT\sim e^D$) till some threshold $D^*< 0.6$.  Beyond this certain distance ($D>D^*$),  all the trajectories take significant small time to reach to the surface of the circle having radius ($D \approx D^*$). In effect, $TT$ saturates around approximately $1800$ for  the current choices of parameters. So, for $D<D^*$, $TT$ has an exponential growth and and beyond, it saturates to a specific value. This essentially tells that no matter where one starts in the basin, the maximum $TT$ that could be achieved is approximately similar (with some small fluctuations) to that of starting from $D^*$. Thus, the probability density function of the transient time is bounded from above by this maximum value of $TT$.  %Therefore, the probability of $TT$ at high value of $TT$  will be higher is  confirmed in the main article.\\

\subsection{Transient time for $\M_2$}
In $\M_2$, we take the size of  basin of attraction to be $[-20,20]\times[-20,20]\times[0,30]$. Performing a similar analysis as above for the averaging, we have plotted the histogram for $TT$ in Fig.\ 2g in the main text. Here, we find that $P(TT)$ is an exponential distribution, which is a fingerprint of chaotic systems \cite{yorkeprl1986}. 

\begin{figure}[t]
	\centerline{
		\includegraphics[scale=0.25]{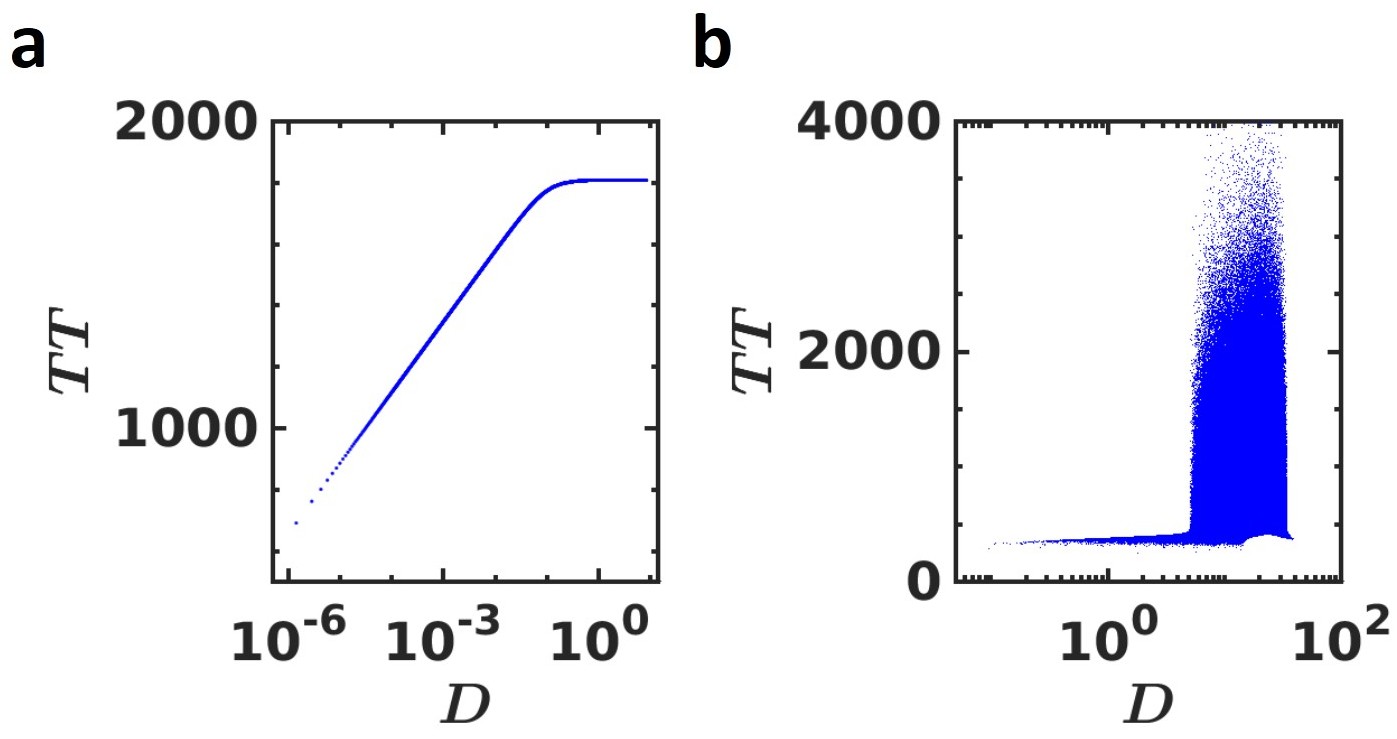}}
	\caption{ Variation between $D$ and $TT$: Transient time as a function of $D$, the distance between the initial point and the target for $\M_1$ (panel a) and $\M_2$ (panel b). For $\M_1$ (panel a), we find that $TT$ increases exponentially as a function of $D$ till it reaches a threshold and then saturates. The threshold value for $D$ is estimated to be $\sim 0.6$. On the other hand, it is clear from panel b ($M_2$) that there is no such relationship between $TT$ and the distance $D$. Parameter values set for the simulations are for (a) $a=-0.01, \Omega=1$, and for (b)  $\sigma=10,\rho=23,\beta=\dfrac{8}{3}$}
	\label{figs1} 
\end{figure}

However, we did not find any direct relationship between $TT$ and $D$ for the Lorenz system. In higher $D$, $TT$  ranges from low value $500$ to  higher value $4000$. It is clear that $D \gtrsim  5$, the scatter points are dense around $500$-$2500$ (see \fref{figs1}b). The less number of points appear in higher value of $TT$ ($TT \gtrsim  3000$). Therefore, $P(TT)$ is less probable at higher values of $TT$. This information is consistent with the form of  $P(TT)$ [see Fig. 2g in the main text].

\begin{figure}[h]
	\centerline{
		\includegraphics[scale=0.69]{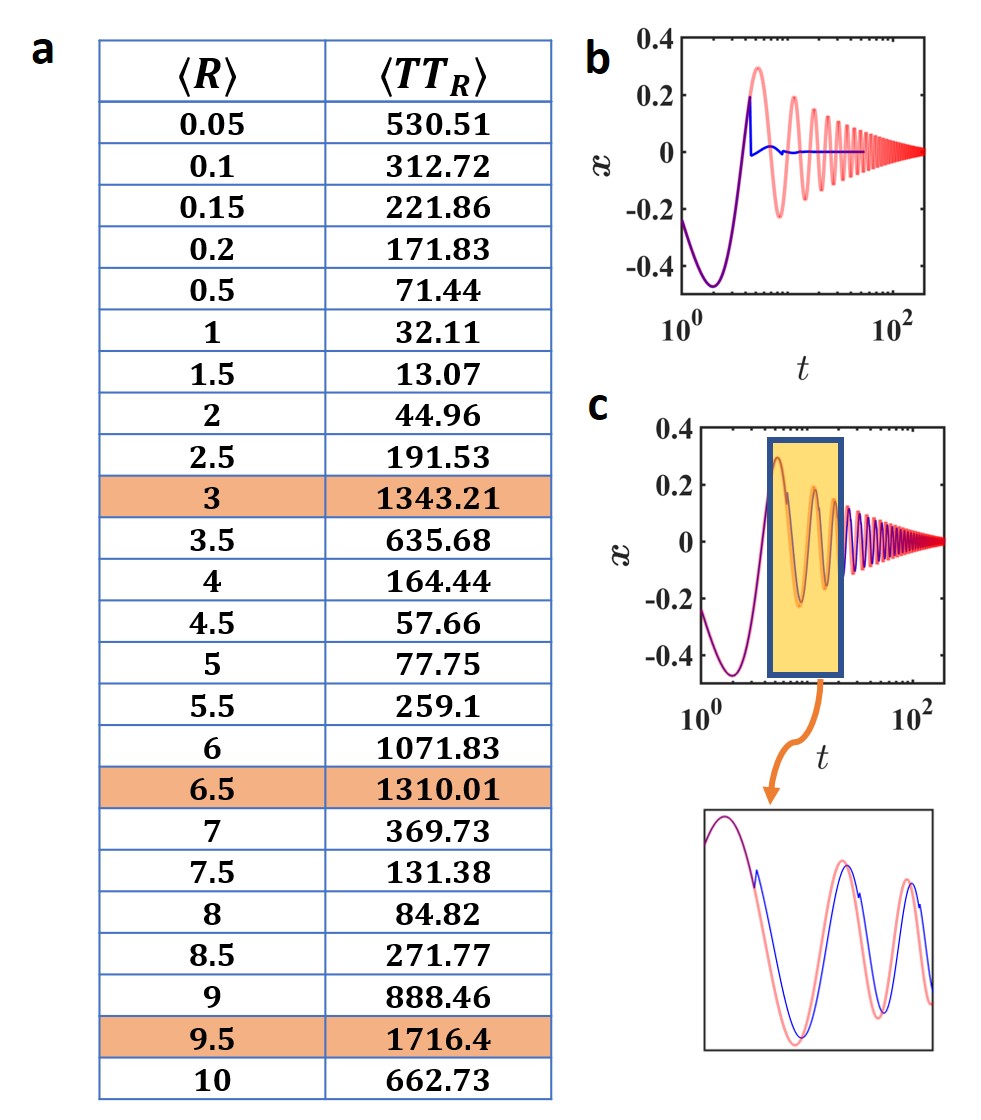}}
	\caption{Panel (a):  Table for mean transient time as a function of $\langle R \rangle$ (in the case of sharp resetting) for model $\M_1$. Marked in red are the rows for which the period $\langle R \rangle$ of sharp resetting  is approximately close to the half or integer function of the intrinsic time period ($T$) of the original process. Time series (trajectory in $x$-coordinate as a function of time) without (in red) and with (in blue) resetting: in panel (b), we have plotted the trajectory for $\langle R \rangle=4.5$ against the original trajectory. We see a clear distinction between the original and resetting induced trajectories. In particular, the plot shows that the trajectory with resetting reaches the target much faster than the original one thus resulting in a lower $\langle TT_R \rangle$. In panel (c), we have plotted the trajectories when  $\langle R \rangle=6.5$ (recall $T \approx 6.3$). We see that the trajectories almost follow each other (also see the inset where we have zoomed a part of both the signals) clearly indicating that both take almost same time to reach the target. Thus, in this case, the behavior of the resetting trajectory is clearly oscillatory like the original process. In other words, resetting will have almost no effect on the underlying process. Parameter values: set here are $a=-0.01, \Omega=1$.
		%	and time series under deterministic resetting:  Time signals before resetting (red) and after resetting (blue) are presented where (b) 		$\langle TT_R \rangle$ attains minimum value at 		$\langle R \rangle=4.5$		whereas for (c) 		$\langle TT_R \rangle$ 		attains considerably higher value at $\langle R \rangle=6.5$. We have zoomed a same part of both signals (blue box with yellow patch). Both trajectories are taking almost same time to converge.
	}
	\label{figs5} 
\end{figure}

\section{Emergence of oscillatory behavior under sharp resetting in $\M_1$}
In this section, we briefly discuss the origin of the oscillatory behavior of $\langle TT_R \rangle$ under sharp resetting mechanism in $\M_1$. This protocol essentially asserts that one resets the system always after a fixed $\langle R \rangle$ amount of time. Note that this oscillatory behavior is markedly different than the exponential resetting where we observed a simple non-monotonic behavior (Fig. 3 in the main text). To explain this, at first, we accumulated $\langle TT_R \rangle$ for different values of $\langle R \rangle$ shown in Fig.\ \ref{figs5}a (Table). Moreover, we recall from Sec. \ref{Eigen_M_1} that the intrinsic periodicity of $\M_1$ model is around  
$T=\frac{2\pi}{\Omega} \approx 6.3$ for $\Omega=1$ (Eq.\ \ref{T_M1}). We now identify from the table (Fig.\ \ref{figs5}a) the \textit{light red marked rows} that satisfy
\bea
\langle R \rangle \approx \dfrac{nT}{2}, n=1,2,3,...,
\eea
where $T$ is the intrinsic period. From the red marked rows of the table, we identify the mean resetting time $\langle R \rangle: 3,6.5,9.5$ for which we respectively find $\langle TT_R \rangle\approx(1343.21, 1310.01, 1716.4)$, which are notably much higher than the transient time one would expect under resetting. 
Essentially, these mean resetting times commensurate with the intrinsic time periods and we observe a significant increase in $ \langle TT_R\rangle $. To further illustrate this behavior, we now choose  two particular values of  $ \langle R \rangle$ from the table such that one lowers the transient time while the second one does not provide any significant improvement. At first, we take $ \langle R \rangle=4.5$ 
which reduces the transient time (In Fig.\ \ref{figs5}b, blue line indicates time signal under sharp resetting which is placed in contrast to the original time series in the absence of resetting). Here, we clearly see a very quick convergence to the steady state for the trajectory subject to resetting. On the other hand, when 
$\langle R \rangle=6.5$,  Fig. \ref{figs5}c clearly indicates that the blue line (which is the trajectory under resetting) is quite close to the original time signal (denoted by red solid line). A short segment of the signal is zoomed in below Fig. \ref{figs5}c to further demonstrate the proximity between the trajectories. Thus, in effect, the resultant transient time becomes of the same order as that of the uninterrupted process. In summary, sharp restarts are periodic temporal process which occur always after a fixed time $\langle R \rangle$. When this period becomes half or full integer of the intrinsic time period of the system, a sudden rise in mean transient time is observed with the emergence of those consecutive oscillations as seen in Fig. 3 (left panel for $\M_1$) in the main text.

\begin{figure}[t]%[H]
	\centerline{
		\includegraphics[width=14.5cm, height=11cm]{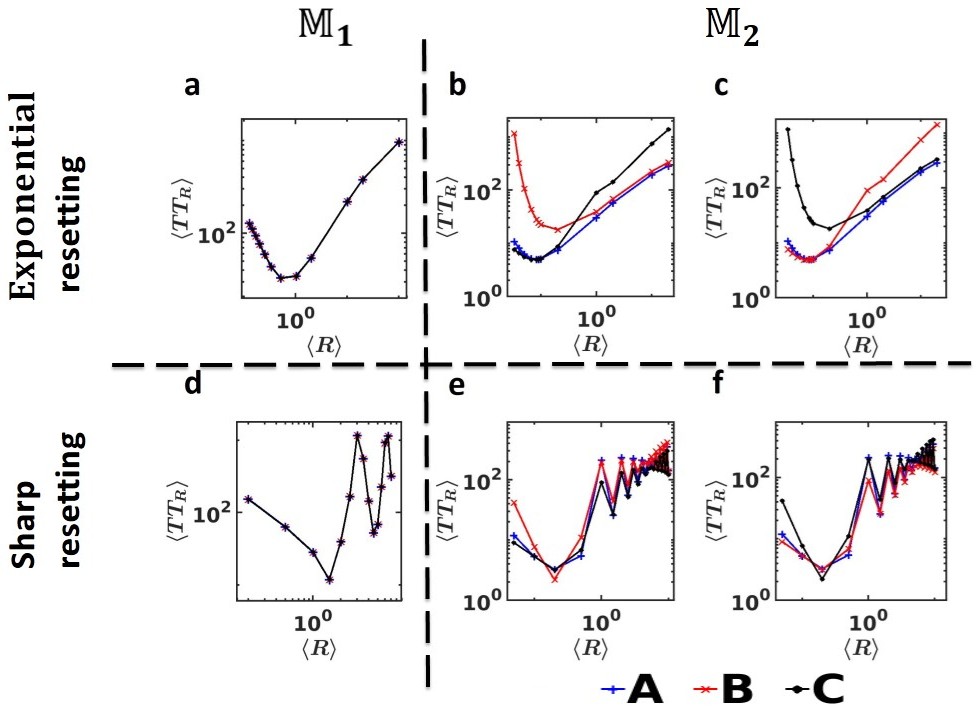}}
	\caption{Variation in mean transient time for different control lines. We have chosen different type of control lines as mentioned in details in Sec.\ V. For $\M_1$, we have considered three different control lines which pass through the points $A : (4,0), B: (-2,4), C: (-4,-2)$ and the equilibrium point $P:(0,0)$ respectively.  For each of these cases, we have plotted the mean transient time as a function of $\langle R \rangle$ [panel (a) for exponential and panel (d) for sharp]. We find that there is no effect of different control lines on the mean transient time in $\M_1$. In $\M_2$, there are two equilibrium points $P_1: (7.65942, 7.65942, 22)$ and $P_2: (-7.65942, -7.65942, 22)$. We have also taken three points $A: (0,0,0)$,  $B:(20,20,30)$, and $ C:(-20,-20,30)$ through which control lines pass. Thus, there are two sets of control lines each of which comprises three lines passing through $A, B, C$ and either $P_1$ or $P_2$. In panel (b) and panel (e), we have plotted $\langle TT_R \rangle$ as a function of $\langle R \rangle$ for exponential and sharp resetting respectively using the control lines that pass through $A, B, C$ and $P_1$. We have prepared similar plots in panel (c) and panel (f) where the control lines pass through $A, B, C$ and $P_2$. Here, we see that mean transient time depends on the choice of control lines. This is due to the nature of the basin for the Lorenz system as discussed in details in Sec.\ VB. Parameter values set here are: $a=-0.01, \Omega=1$ (for $\M_{1}$) and  $\sigma=10,\rho=23,\beta=\dfrac{8}{3}$ (for $\M_{2}$).}
	\label{figs6} 
\end{figure}

\begin{figure}[t]%[H]
	\centerline{
		\includegraphics[width=10.5cm, height=9cm]{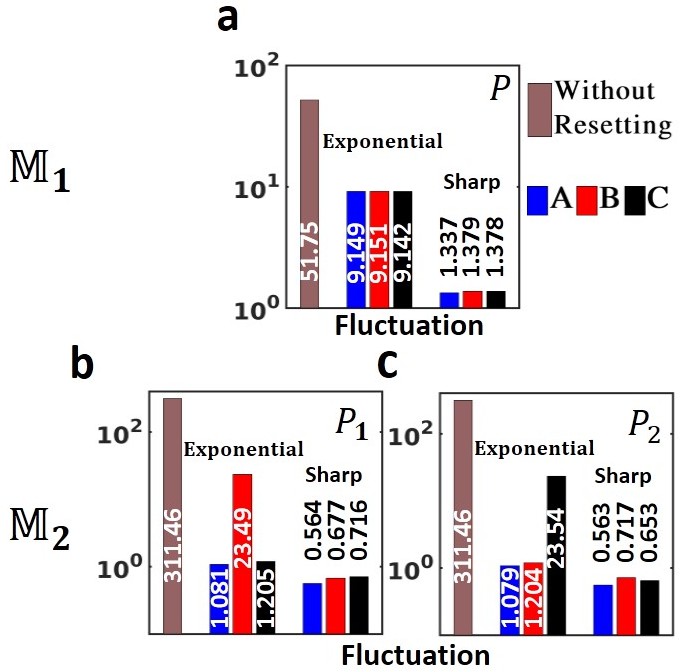}}
	\caption{ Variation in fluctuations for different choices of control lines. In panel (a), we have shown a bar plot comparison of fluctuations between the original dynamics and resetting induced dynamics in $\M_1$. Resetting was conducted at $\langle R \rangle=1$ (both for exponential and sharp resetting) by taking the control lines which pass through $P$ and $A, B,C $ respectively (see Sec.\ VA). It is clear from the figure that (i) resetting reduces fluctuations and (ii) the magnitude of the fluctuations is almost same implying that resetting does not depend on the choice of control lines in $\M_1$. This observation is in accordance with Fig.\ 6a and Fig.\ 6d. Panel (b) and panel (c) show bar plot comparison of fluctuations between the original dynamics and resetting induced dynamics in $\M_2$ (conducted at $\langle R \rangle=0.1$) when the control lines pass through $A, B,C $ and either $P_1$ or $P_2$ respectively (see Sec.\ VB). We concur with the observation that resetting also reduces fluctuations in this case. However, the magnitudes of fluctuations are different in each case, as expected, due to the underlying non-uniform structure of the basin in Lorenz system. Parameter values set here are: $a=-0.01, \Omega=1$ (for $\M_{1}$) and  $\sigma=10,\rho=23,\beta=\dfrac{8}{3}$ (for $\M_{2}$).}
	%	Variation of $\langle TT_R \rangle$  with respect to $\langle R \rangle$ for chosen three lines, passing through A(blue), B(red), C(black) and $P$ for $\M_1$ system,  $P_1$ and $P_2$ for $\M_2$ system under (a,b,c) stochastic resetting and (d,e,f) deterministic resetting. The chosen points are A: $(4,0)$, B: $(-2,4)$, and C: $(-4,-2)$ for Stuart-Landau system and A: $(0,0,0)$, B: $(20,20,30)$, and C: $(-20,-20,30)$ for Lorenz system. {\color{blue} The first, second and third columns represent the results for the lines passing through (a, d) $P$ for $\M_1$ system, (b, e) $P_1$ and (c, f) $P_2$ for $\M_2$ system, respectively. } }
	\label{fluctuations} 
\end{figure}

\section{Behavior of the mean and fluctuations in transient time on the choice of control lines}
In this section, we investigate in details the ramifications in $\langle TT_{R} \rangle$ and fluctuations $\sigma_R$ on the choice of control lines. Let us first recall that a control line is randomly chosen from the basin of attraction but it always passes through the equilibrium point(s). Here, the analysis is done both for the exponential and sharp resetting. In the following, we discuss the effects of control line on the mean and fluctuations first on the Stuart-Landau system ($\M_1$), and then on the Lorenz system ($\M_2$).

\subsection{Effect of control lines on $\M_1$}
In system $\M_1$, the equilibrium point is located at $(0,0)$ which we denote as $P$. In the main text, we choose the control line randomly from the basin such that it passes through $(4,4)$ and the equilibrium point $P$. We have shown that this protocol yields a significant reduction in mean and fluctuations in transient time. To show that this behavior is invariant to the choice of the control line, we now construct the following control lines which pass through the random coordinates mentioned below from the basin of attraction:
\begin{enumerate}
	\item  $P (0,0)$ and  $A(4,0)$ ,
	\item  $P (0,0)$ and  $B(-2,4)$ , 
	\item  $P (0,0)$ and  $C(-4,-2)$.
\end{enumerate}
For each of the cases above, we have plotted $\langle TT_R \rangle$ as a function of $\langle R \rangle$
for the exponential (Fig.\ \ref{figs6}a) and sharp resetting (Fig.\ \ref{figs6}d) respectively. First, we note that indeed resetting reduces the mean transient time. Secondly, it becomes evident from the plots that all the curves collapse thus clearly indicating the fact that the variation in mean transient time does not depend on the choice of the control line, particularly, for case of $\M_1$, where the basin of attraction is homogeneous, and thus the system can not  distinguish between the choice of the control lines. In Fig. \ref{fluctuations}a, we have shown a comparison between the fluctuations in the original dynamics and with resetting dynamics (both for exponential and sharp) for given $\langle R \rangle =1$. Note that the fluctuations are now reduced due to the resetting. Moreover, since the basin is uniform, the choice of control line did not have any impact on the fluctuations similar to the mean as seen above.

\subsection{Effect of control lines on $\M_2$}
To see the effects of control lines on $\M_2$, we first recall that $\M_2$ has two fixed points ($P_1$ ~and~ $P_2$) which are stable for a certain range of $\rho$ (See the Sec.  \ref{Eig_M2}).  The system $\M_2$ has riddle basin of attraction for the equilibrium points $P_1$ and $P_2$. As was mentioned in the main text, in this case, we have some flexibility in choosing control lines e.g., it can pass through one of the equilibrium points ($P_1$ or $P_2$) or via both. We discuss each of the cases in the following.

\subsubsection{Effect of fixed control line passing through both $P_1$ and $P_2$}
We first discuss the case when the control line passes through both the equilibrium points $P_1$ and $P_2$. We compute the transient time when the trajectory reaches any of these points. This scenario was already discussed in the main text. In particular, we choose the control line such that it passes through  $P_1 (7.65942, 7.65942, 22)$ and  $P_2(-7.65942,-7.65942, 22)$. When conducted at $\langle R \rangle=0.1$, a net reduction in mean and fluctuations was observed.

\subsubsection{Effect of fixed control line passing through $P_1$}
In this case, we choose control lines that pass through the equilibrium point $P_1$, which is the only target. Here, we take three random control lines passing through the following points from the basin of attraction as described below
\begin{enumerate}
	\item $P_{1}$   $(7.65942, 7.65942, 22)$ and  $ A(0,0,0)$,
	\item $P_{1}$   $(7.65942, 7.65942, 22)$ and  $B(20,20,30)$,
	\item $P_{1}$   $(7.65942, 7.65942, 22)$ and  $C(-20,-20,30)$.
\end{enumerate}
In Figs.\ \ref{figs6}b and \ref{figs6}e, we have plotted $\langle TT_R \rangle$ as a function of $\langle R \rangle$ for the exponential and deterministic resetting respectively. The behavior is similar to Fig. 3 in the main text which essentially reiterates the fact that resetting reduces the mean transient time. In Fig. \ref{fluctuations}b, we have shown a comparison between the fluctuations in the original dynamics and with resetting dynamics (both for exponential and sharp) for given $\langle R \rangle =0.1$ and choice of the control lines as mentioned above. In here, we also see that resetting lowers the fluctuations.

%shown the impact of control lines  for stochastic and deterministic resetting,  respectively. For stochastic resetting, the optimal values generated from black and blue lines are better than the choice of red line. {\color{blue} This is because of the red line passing through the two points $P_1 (7.65942, 7.65942, 22)$ and  $B(20,20,30)$ which lie in the positive octant in the phase space. So the initial points  near the other equilibrium point $P_2$ in the basin of attraction take more time to reach $P_1$ through this line.  } 

\subsubsection{Effect of fixed control line passing through $P_2$}
In this case we take the control line passing through the equilibrium point $P_2$ (which is the only target) and the following other points
\begin{enumerate}
	\item $P_{2}$   $(-7.65942, -7.65942, 22)$ and  $ A(0,0,0)$,
	\item $P_{2}$   $(-7.65942, -7.65942, 22)$ and  $B(20,20,30)$,
	\item $P_{2}$   $(-7.65942, -7.65942, 22)$ and  $C(-20,-20,30)$.
\end{enumerate}
Here too, we find that resetting using a control line technique reduces the mean transient time. These conclusions are in accordance with the  Figs.\ \ref{figs6}c and \ref{figs6}f which show the variation of mean transient time as a function of $\langle R \rangle$. In Fig. \ref{fluctuations}c, we have shown a comparison between the fluctuations in the original dynamics and with resetting dynamics (both for exponential and sharp) for given $\langle R \rangle =0.1$ and choice of the control lines as mentioned above. In here, we also see that resetting lessens the fluctuations. 
\\

\noindent
As a final remark, we note that since the basin of $\M_2$ is non-homogeneous, we do not observe any collapse for the mean transient time for different choices of control lines as was seen in the case of $\M_1$.

%{\color{blue}Here, blue and red control lines outperform the black line. So for this bistable system ($\M_2$), any arbitrary lines passing through the octants containing the two equilibrium points $P_1$ and $P_2$ give the same effect under stochastic resetting process.   {\color{red} \bf We can delete this line now. In future, we would to like explore these discrepancies  (issues) in details.} Note that, the different choices of  control lines do not have separate impacts in case for deterministic resetting.}

% For model $\M_2$, two cases are arise in such a way that variation of $\langle TT_R \rangle$ with respect to $\langle R \rangle$ by choosing the straight lines are passing through $FP_1$ and A: $(0,0,0)$, B: $(20,20,30)$, and C: $(-20,-20,30)$ under stochastic and deterministic strategy in Fig.\ \ref{fig4}(b) and Fig.\ \ref{fig4}(e) and $FP_2$ and A: $(0,0,0)$, B: $(20,20,30)$, and C: $(-20,-20,30)$ under stochastic and deterministic strategy in Fig.\ \ref{fig4}(c) and Fig.\ \ref{fig4}(f), respectively.   
\section{Impact of  control parameters on the transient time near the critical transition}\label{parameter}
\noindent
It is well known that in non-linear systems, the parameters play a paramount role to decide the structure of the basin, attractor or fixed points. For our current models, we have already discussed in Sec.\ \ref{Stability} that the controlling parameters can change the structure of the attractor qualitatively i.e., transform the stable fixed points into
limit cycle or chaos (beyond critical values). But it is important to note that as the parameters are tuned to the critical values, duration of transient states gradually increases. For example, it is known that the average lifetime or transient time of a chaotic transient depends critically upon the system parameter i.e., it diverges as a power law form near the critical point \cite{yorkeprl1986}. Naturally, the question appears on how the situation changes in the presence of resetting near the critical point and what are the overall ramifications of the resetting strategies (exponential and sharp) on the statistics of the transient time. In this section, we have examined these issues in details.

\begin{figure}[t]
	\centerline{
		\includegraphics[height=10cm, width=15cm]{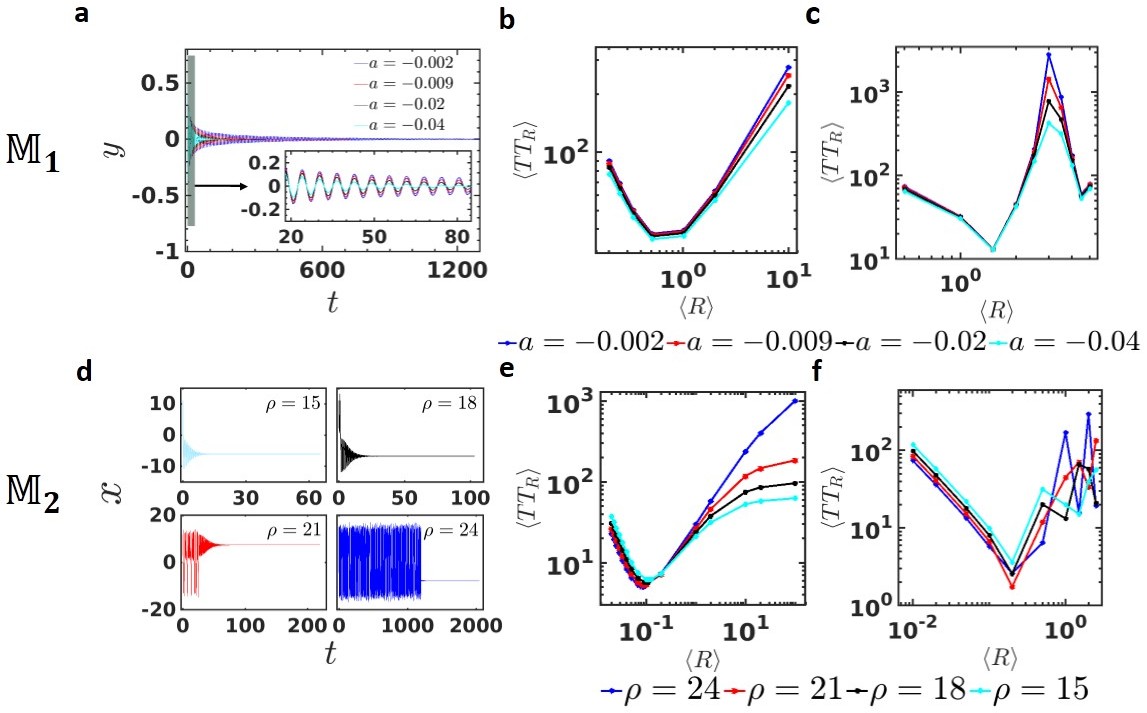}}
	\caption{Mean transient time regulation by resetting near the critical transition. Panel (a) and panel (d) show four time series of the original dynamics for $\M_1$ and $\M_2$ respectively. In panel (a), the trajectories in coordinate $y$ are plotted as a function of time for different values of $a$ (shown in the plot) near the critical transition $a_c=0$. In panel (d), the trajectories in coordinate $x$ are plotted as a function of time for different values of $\rho$ (shown in the plot) near the critical transition $\rho_c=24.06$. The varying parameters are  $a=-0.002$ (blue), $-0.009$ (red), $-0.02$ (black), $-0.04$ (cyan), and $\rho=15$ (cyan), $18$ (black), $21$ (red), $24$ (blue).  In panel (b) and (c), we have plotted  $\langle TT_R \rangle$ as a function of $\langle R \rangle$ for exponential and sharp resetting for the above mentioned values of $a$. Similarly, panel (e) and panel (f) depict variation of $\langle TT_R \rangle$ as a function of $\langle R \rangle$ for exponential and sharp resetting for the above mentioned values of $\rho$. 
		Other parameters set for the simulations are: $\Omega=1$ (for $\M_1$), and $\sigma=10$, $\beta=\dfrac{8}{3}$ (for $\M_2$).}
	%	\caption{\textbf{Time series of the original dynamics for different values of parameters:} Four different time series for four values of $a$ for (a) $\M_1$ and $\rho$ for (e) $\M_2$. We scan $a$ and $\rho$ and approach towards their  critical values $a_c$ and $\rho_c$ so that the systems undergo critical transitions.	\textbf{Mean transient time in presence of  resetting :} panel (a) depicts variation of $\langle TT_R \rangle$ as a function of $\langle R \rangle$  four values of $a$ for (b) $\M_1$ and four values of $\rho$ for (f) $\M_2$ under exponential resetting.		\textbf{Plot of PDF of transient time:} PDF of transient time $TT_R$ at $\langle R^* \rangle$ for the systems (c)  		$\M_1$ and (g) $\M_2$. 		\textbf{Mean transient time under deterministic resetting process:} $\langle TT_R \rangle$ versus $\langle R \rangle$ in the log-log scale for four values of $a$ for (d) $\M_1$ and for four values of $\rho$ for (h) $\M_2$ under deterministic resetting. The varying parameters are  $a=-0.002$ (blue), $-0.009$ (red), $-0.02$ (black), $-0.04$ (cyan), and $\rho=15$ (cyan), $18$ (black), $21$ (red), $24$ (blue). Other parameters: $\Omega=1$, $\sigma=10$, $\beta=\dfrac{8}{3}$.}
	\label{fig5} 
\end{figure} 

\subsection{System $\M_1$}
In the Stuart-Landau oscillatory system, we regulate the decay parameter $a$ which determines whether the system has a limit cycle or a fixed point. Following analysis from Sec. \ref{Stability}A, we know that this transition occurs exactly at $a_c=0$. In what follows, we scan $a$ for a range of values close to $a_c$ and examine the variations due to resetting. For a  given initial condition, the transient time of the underlying process gradually increases as we increase $a$. This is shown in
Fig. \ref{fig5}a where $a$ has assumed four different values $-0.002, -0.009, -0.02, -0.04$ and clearly, as  $|a|$ increases the decay rate of the oscillation increases and we see a faster convergence (i.e., a shorter transient time) to the steady state (see inset in Fig. \ref{fig5}a). To add restart, we follow the same protocol (by resetting at the control line that passes through the equilibrium point $P$) as outlined in the main text to this dynamics but when $a$ is close to $a_c$. In Fig. \ref{fig5}b, we have plotted $\langle TT_R \rangle$ as a function of $\langle R \rangle$ [$a=-0.002$ (blue), $-0.009$ (red), $-0.02$ (black), and $-0.04$ (cyan)] when the resetting is exponential. The plot clearly shows that $\langle TT_R \rangle$ is significantly reduced near the critical transition. Moreover, in each case above, we find an optimal resetting time $\langle R^* \rangle$ which makes $\langle TT_R \rangle$ to be minimum (see Table I for exponential and Table II for sharp resetting and details of the mean transient time at the optimality). We prepare a plot in Fig. \ref{fig5}c for the sharp resetting case where we find behavior of  $\langle TT_R \rangle$ to be similar. The oscillatory behavior, as was discussed in Sec. IV, was noted for the sharp resetting.

%There is no significant differences among the average $TT_R$  for  $a=-0.002$ (blue), $-0.009$ (red), $-0.02$ (black), and $-0.04$ (cyan) ($\M_1$,  Fig.\ \ref{fig5}b).   Optimal value of $\langle R \rangle$ has not been  changed, and not only that, the value $\langle TT_R \rangle$ at $\langle R^*\rangle=0.5$ is almost same for each $a$.

\subsection{System $\M_2$}
In the Lorenz system, it is known that the Rayleigh number $\rho$ marks the critical transition between the chaotic transient phase and chaotic attractor \cite{yorkeprl1986,yorke1979}. For fixed parameters $\sigma=10$ and $\rho=\frac{8}{3}$ , the system exhibits transient chaos in the range of $\rho \in (1.926,24.06)$ where the transition to a chaotic attractor takes place at $\rho_c=24.06$. To demonstrate the effects of resetting near the critical transition, we take four different values for $\rho$ and plot the trajectories for each of them. We demonstrate in Fig.\ \ref{fig5}d, trajectories in $x$-coordinates as a function of time for $\rho=24$ (blue), $21$ (red), $18$ (black), and $15$ (cyan). Here, chaotic transient phase persists longer as we increase $\rho$ close to $\rho_c$. To illustrate the effects of resetting, we plot $\langle TT_R \rangle$ as a function of $\langle R \rangle$ for each of the cases above (by taking a control line which passes through both the equilibrium points $P_1$ and $P_2$). Both for exponential (Fig.\ \ref{fig5}e) and sharp resetting (Fig.\ \ref{fig5}f), we observe that resetting reduces the transient time which would be significantly higher and even diverging (close to $\rho_c$). Moreover, emergence of an optimal resetting rate $\langle R^* \rangle$ was observed in each case (see Table I for exponential and Table II for sharp resetting and details of the mean transient time at the optimality).

%The same type of feature is confirmed for $\M_2$ at     $\rho=24$ (blue), $21$ (red), $18$ (black), and $15$ (cyan) ($\M_2$,  Fig.\ \ref{fig5}f). In this  case,  optimal $\langle R\rangle$ i.e. $\langle R^{*}\rangle$ is sightly shifted. $\langle R^*\rangle$ increases if we increase $\rho$, i.e., $\langle R^*\rangle=0.11,0.1,0.091,$ and $0.083$ for $\rho=15,18,21$, and $24$, respectively. 

%the higher values of $\rho$ and $a$ unable to reduce the transient time. These are the cases which occur naturally, i.e., in absence of resetting.

% \par Probability densities of $TT_R$ at $\langle R^* \rangle$ corresponding to those aforementioned parameters of $\M_1$ and $\M_2$ are depicted in Figs.\ \ref{fig5}c and \ref{fig5}g, respectively. Here, shifting of $\langle TT_R \rangle$ with changing of parameter value is clearly exhibited.   Similarly, for deterministic resetting, $\langle R \rangle$ versus $\langle TT_R \rangle$ is also delineated for $\M_1$ in Fig.\ \ref{fig5}d and for $\M_2$ in Fig.\ \ref{fig5}h. Here, we have observed $\langle R^* \rangle=1.5$ for system $\M_1$ and $\langle R^* \rangle =0.2$ for system $\M_2$. 

%   Below we give a chart for stochastic resetting which summarize the approximated values of $\langle TT_R^* \rangle$ attains at $\langle R^* \rangle$ for all considered parameters in Table I.         
\begin{center}
	Table I
	\begin{tabular}{ |c|c|c|c| } 
		\hline
		$a$ & $<TT_R^*>$ & $\rho$ & $<TT_R^*>$\\
		\hline
		{$-0.04$}  & 35.40 & {$15$}  & 6.16 \\ 
		\hline
		{$-0.02$}  & 36.63 & {$18$}  & 5.70 \\ 
		\hline
		{$-0.009$} & 37.34 & {$21$}  & 5.30 \\ 
		\hline
		{$-0.002$} & 37.81 & {$24$}  & 5.05\\ 
		\hline
		
	\end{tabular}
\end{center}

%Similarly, another chart for deterministic resetting, the approximated values of $\langle TT_R^* \rangle$ at optimal $\langle R^* \rangle$ are provided for all considered parameters in Table II.         

\begin{center} 
	Table II
	\begin{tabular}{ |c|c|c|c| } 
		\hline
		$a$ & $<TT_R^*>$ & $\rho$ & $<TT_R^*>$\\
		\hline
		{$-0.04$} & 12.905 & {$15$}  
		& 3.59\\ 
		\hline
		{$-0.02$} & 13.02 & {$18$}  
		& 2.55\\ 
		\hline
		{$-0.009$} & 13.07 & {$21$}  
		& 1.74\\ 
		\hline
		{$-0.002$} & 13.10 & {$24$}  
		& 2.59\\ 
		\hline
	\end{tabular}
\end{center}

\vspace{1cm}

\begin{figure}[t]
	\centerline{
		\includegraphics[height=8cm, width=12cm]{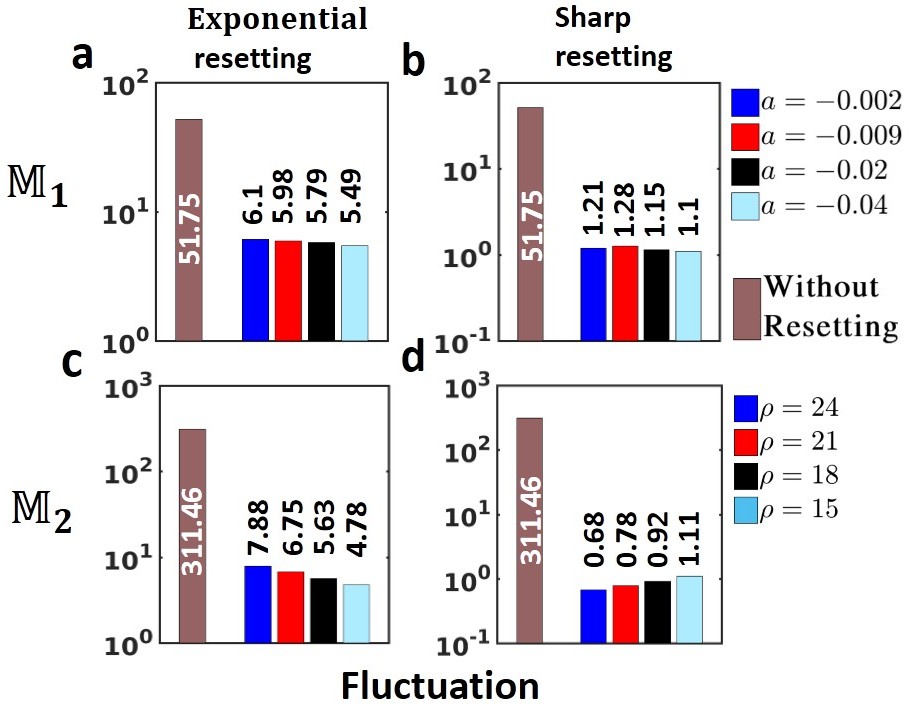}}
	\caption{Fluctuation regulation by resetting near the critical transition. In this figure, we present bar plot comparison between the fluctuations of the original and reset induced dynamics. For $\M_1$, while conducted at $\langle R \rangle=1$, we observe that both exponential (panel a) and sharp (panel b) resetting strategies have reduced the fluctuations even when we are close to the critical transition $a=a_c=0$. Similar bar plot is laid out for $\M_2$ but resetting here was conducted at $\langle R \rangle=0.1$. Here too, we find that resetting remains beneficial to reduce fluctuations as we scan $\rho$ to its critical value $\rho_c=24.06$. }
	\label{fig6} 
\end{figure}

\noindent
Finally, we conclude this section by reemphasizing the fact that resetting has a strong impact on the average transient times even close to the critical transition. In particular, resetting renders the mean transient time lower near the critical point which are otherwise large or diverging. It is worth emphasizing that resetting also regulates the fluctuations strongly near the critical transition. We refer to the barplot in \fref{fig6} which clearly shows that there is a significant reduction in fluctuations even when we modulate the parameters very close to the critical transition. A consistent limit is obtained for $\langle R \rangle \geq10$, where the system behaves as it would in the absence of resetting.

\section{Computational method}\label{computational}
\noindent
In this section, we briefly discuss the computational method that has been used to gather statistics and perform averaging on the transient time under exponential (stochastic) and sharp (deterministic) resetting strategy.
\begin{itemize}
	\item {\bf {\it Step I.} Fix the target:} 
	First, we determine the equilibrium point $\mathcal{A}$ of the given differential equation. There can be many equilibrium points in the system, but we may choose one or many of them to be the target points. For brevity, let us denote the specific targeted fixed point by $\mathbf{x_f}$. 
	%At first, we find out whether the given system possesses the equilibrium point or not. If it possesses, then how many stable equilibrium points are by linear stability analysis. Let $\bf x_f$ be equilibrium point of the given differential equation.\\     
	\item {\bf {\it Step II.} Integration scheme:}  To integrate the deterministic model,  we choose a random initial condition, say $\bf x_{0}$ from the basin of attraction $\mathcal{B_A}$ at the initial time $T_0$. The $4$-th order Runge{-}Kutta method is used to simulate the system with  fixed  step length $h=0.01$.  Sufficient number of data points are generated such that  trajectory reaches to its target with a close vicinity measured by $\epsilon=10^{-9}$ i.e., it satisfies Eq.\ \ref{tt} in the main text. %We have checked that the phenomena do not change if we slightly change the value of $h$.  
	\item {\bf {\it Step III.} Generating resetting times:}
	Starting from $T_0$, we now evolve the dynamics under resetting mechanism. Resetting events occur at time $T_{1}, T_{2}, T_{3},...$, where the duration between two consecutive events ( $\Delta_T: \{T_1-T_0, T_2-T_1, T_3-T_2,...\}$) are extracted from an exponential distribution
	\begin{align}
		f_R(\Delta_T)=\langle R \rangle ^{-1} e^{- \frac{\Delta_T}{\langle R \rangle}},~~~\text{where}~\langle R \rangle~~\text{is the mean}
	\end{align}
	and periodic distribution for sharp resetting 
	\begin{align}
		f_R(\Delta_T)=\delta(\Delta_T-\langle R \rangle),~~~\text{where}~\langle R \rangle~~\text{is the fixed time period}~.
	\end{align}
	For numerical schemes, the resetting times  were generated  at the discrete points: $\frac{1}{h}\times  \{T_{1}, T_{2}, T_{3},...\}$. 
	
	%In our paper, for stochastic resetting, the duration  between two consecutive events ( $\Delta_T: \{T_1-T_0, T_2-T_1, T_3-T_2,...\}$) are extracted from the exponential distribution and the corresponding density function looks like  $f_R(\Delta_T)=\langle R \rangle ^{-1} e^{- \frac{\Delta_T}{\langle R \rangle}}$. Instead of probabilistic resetting, the times for resetting the trajectory can be collected in a deterministic way {\it i.e.} $\Delta_T$=constant. Corresponding density function can be written as $f_R(\Delta_T)=\delta(\Delta_T-\langle R \rangle$). %For both cases, $\langle R \rangle$ must be fixed %previously.
	
	\item  {\bf {\it Step IV.} Fixing a control line:} An arbitrary point $\bf x_{c}$ is randomly chosen from the basin of attraction and we draw a straight line passing through $\bf x_{c}$ and any of the equilibrium point(s), say, $\bf x_{f}$. This arbitrary control line is kept fixed for the entire scanning process. We have scanned the transient times of $5 \times 10^6$  initial states  for each $\langle R \rangle$. 
	%The corresponding basin of the attraction(s) of equilibrium point(s) must be figured out. If the system becomes monostable, then a preassigned control line passing through the equilibrium point is fixed in the basin of attraction. But, if it shows more than one stable equilibrium point, then choices are up to us irrespective of the structure of basin of attraction (well separated or riddled). For example, for the case of bistability, if we want the transient time for the system, then we chose the control line passing through either anyone of equilibrium points or both. Remember one thing that the arbitrary control line is remain same through out the process after fixing once.  {\color{red} What happened if limit cycle co-exists with equilibrium point} 
	
	\item {\bf {\it Step V.} Projection procedure:} %The most important part is how to reset a trajectory when the clock indicates for starting the process at any instant time.
	To describe the projection or resetting to the control line, let us first assume that resetting occurred at some time $T_i$, and at this very moment, coordinate of the trajectory is ${\bf x_1}$. To decide, where to reset in the control line, we choose a point $\bf{x_2}$ from the control line such that the line passing through ${\bf x_1}$ and ${\bf x_2}$ will be perpendicular to the control line. If this condition is satisfied, we project the coordinate ${\bf x_1}$ to ${\bf x_2}$. This process is repeated for other resetting events.
	%Let us assume that, the resetting clock shows the time $T_1$ and   the position of the trajectory be ${\bf x_1}$. In this time, the  current position ${\bf x_1}$ will be  reset at  ${\bf x_2}$. Note that, ${\bf x_2}$ is carefully chosen from the  control line such  that, the  control line and the line passing through ${\bf x_1}$ and ${\bf x_2}$ will be perpendicular to each other. We continue the same process for the rest of the resetting times.
	\item {\bf {\it Step VI.} Calculation of  transient time:}   We stop our simulation after reaching at $\bf x_n$ after $n$-th iteration only if the condition $||\bf x_f-\bf x_{n}||<\epsilon~(=10^{-9})$ (See Sec. I and Eq.\ (2) in the main text) is satisfied. Subsequently, the  transient time will be $TT=n\times h$. This time is random, and we generate histogram of the transient time from many such realizations.
	%We already define the transient in the Sec.\ \ref{Notation}. Let, after $n$-th iteration, the position of the trajectory be $\bf x_{n}$. Then if absolute difference between $\bf x_{n}$ and $\bf x_{f}$ be less than pre defined $\epsilon$ (however small real number), then immediately we stop the simulation and consider the transient time, $TT=n\times h$. We fixed the value at $\epsilon=10^{-9}$ in our article. But, changing the value of $\epsilon$, quantitative changes must be changed but qualitative changes remain the same.
\end{itemize}
Following the steps I-VI, we collect data of the required observables and investigate various statistical properties.
\begin{figure}[ht]
	\centerline{
		\includegraphics[scale=0.325]{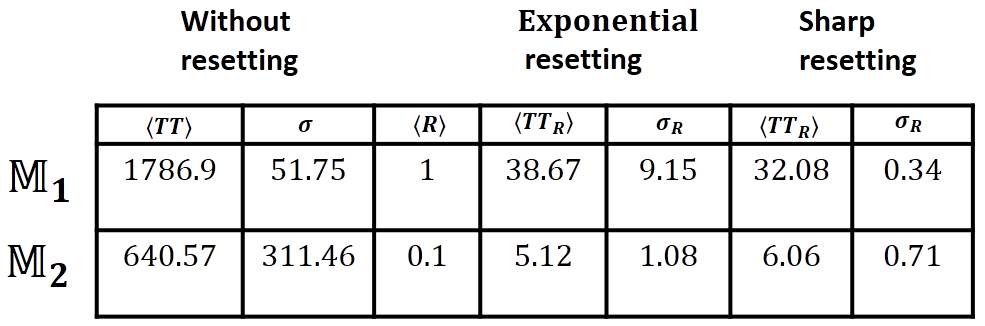}}
	\caption{ Numerical values for the mean and fluctuations as was pointed out in the main text. In $\M_1$, both resetting strategies (exponential and sharp) were conducted at $\langle R \rangle=1$ (also see Fig.\ 2e in the main text for the exponential resetting). In $\M_2$, everything was similar but we took $\langle R \rangle=0.1$ (also see Fig.\ 2j in the main text for the exponential resetting). In both the cases (exponential and sharp), the order of improvement in mean and fluctuations was mentioned in the main text.}
	\label{figtable} 
\end{figure}

\section{Summary of the numerical values used in the main text}
In this section, we provide numerical values for the mean and fluctuations for exponential and sharp resetting as was discussed in the main text. We refer to Fig. \ref{figtable} which contains a table listing the exact values.

\vspace{0cm}

\bibliography{bibliography_maintext-dana_20}
\bibliographystyle{apsrev4-1}

%\begin{thebibliography}{1}
%	
%	\bibitem{Ott}
%	Ott, E., 2002. Chaos in dynamical systems. Cambridge university press.
%	
%	\bibitem{Strogatz}
%	Strogatz, S.H., 2018. Nonlinear dynamics and chaos with student solutions manual: With applications to physics, biology, chemistry, and engineering. CRC press.
%	
%	\bibitem{Yorke} 
%	Yorke, J.A. and Yorke, E.D., 1979. Metastable chaos: The transition to sustained chaotic behavior in the Lorenz model. Journal of Statistical Physics, 21(3), pp.263-277.
%	
%	\bibitem{Grebogi} Grebogi, C., Ott, E. and Yorke, J.A., 1986. Critical exponent of chaotic transients in nonlinear dynamical systems. Physical review letters, 57(11), p.1284.
%\end{thebibliography}	

\end{document}